\documentclass[a4paper,12pt]{article}
\usepackage[british]{babel}
\usepackage{amsfonts}
\usepackage{amsfonts}
\usepackage{amsmath}
\usepackage{amssymb}
\usepackage{graphicx}
\begin{document}

\begin{center}
\baselineskip=1.5
\normalbaselineskip{\Large On a comparative study between
dependence scales determined by linear and non-linear measures}

{\large S\'{\i}lvio M. Duarte Queir\'{o}s}\footnote{%
email address: Silvio.Queiros@unilever.com, sdqueiro@googlemail.com}

\baselineskip=1.0 \normalbaselineskip

\textit{Unilever R\&D Port Sunlight, Quarry Road East, Wirral, CH63 3JW UK \\%
[0pt]
}

\baselineskip=1.0 \normalbaselineskip

{\small (10th December 2008)}
\end{center}

\baselineskip=1.0 \normalbaselineskip

\subsection*{\protect\bigskip Abstract}

In this manuscript we present a comparative study about the determination of
the relaxation (\textit{i.e.}, independence) time scales obtained from the
correlation function, the mutual information, and a criterion based on the
evaluation of a nonextensive generalisation of mutual entropy. Our results
show that, for systems with a small degree of complexity, standard mutual
information and the criterion based on its nonextensive generalisation
provide the same scale, whereas for systems with a higher complex dynamics
the standard mutual information presents a time scale consistently smaller.

\section{Introduction}

The description of the degree of dependence between variables is of capital
importance, namely in several applications like time series analysis in
which it is valuable to define how long there is a relevant relation between
its elements. As examples, we mention: \textit{i)} the determination of time
scales from which value on a system is considered to be in a stationary state,
\textit{i.e.},
\begin{equation}
\frac{\partial P\left( z\left( t\right) ,t\right) }{\partial t}=0,
\end{equation}%
or, in other words, the time needed for a system to achieve such a state ($%
z\left( t\right) $ represents an element of a time series $Z\equiv \left\{
z\left( t\right) \right\} $ at time $t$), \textit{ii)} the existence of
ageing phenomena, \textit{i.e.}, the dependence of the \textit{correlation
function},

\begin{equation}
C_{z}\left( t_{w},\tau \right) =\frac{\left\langle z\left( t_{w}\right)
\,z\left( t_{w}+\tau \right) \right\rangle -\left\langle z\left(
t_{w}\right) \right\rangle \left\langle z\left( t_{w}+\tau \right)
\right\rangle }{\sqrt{\left\langle z\left( t_{w}\right) ^{2}\right\rangle
-\left\langle z\left( t_{w}\right) \right\rangle ^{2}}\sqrt{\left\langle
z\left( t_{w}+\tau \right) ^{2}\right\rangle -\left\langle z\left(
t_{w}+\tau \right) \right\rangle ^{2}}},
\end{equation}%
on the \emph{waiting time}, $t_{w}$, \textit{iii)} the appraisal of how good
the forecasting of future events can be or even how long we can produce
reliable predictions based on previous values of the time series, \textit{iv)%
} embedment of time series used in state space reconstruction and
independent component analysis \cite{swinney,prichard,ling,independbook},
among many other cases.

The most straightforward way of performing this assessment has been the
evaluation of the correlation function. Even though it has widespread
applications, in truth, for a large class of processes, specifically complex
systems~\footnote{%
A complex system has been consensually defined as a system whose behaviour
crucially depends on its details~\cite{parisi}.}, such a procedure is unable
to give a proper answer~\cite{independbook}. Explicitly, the correlation function
is a normalised covariance that is only effective
at determining the dependences which are either linear or can be written in
a linear way. Hence, by simply applying $C_{z}\left( t_{w},\tau \right) $,
the dependences that do not fit in the linear classification can not be
correctly measured. It is worth stressing that non-linear dependences rule a
large part of the systems presently studied~\cite{strogatz}. In other words,
if we aim to characterise this sort of systems we must look at higher-order
correlations to check for statistical independence. In order to make it, the
correlation function is many times replaced by the computation of the mutual
information which is able to detect the existence of non-linearities in the
system \cite{independbook,strogatz,grassberger,grassbergerepl,physrepaus}.
In the present work, we carry out a comparative study between the
correlation function, the mutual information (based on the Kullback-Leibler
entropy~\cite{kullback-leibler-entropy}) and a generalised measure of mutual
information~\cite{ct-kl-1998} which emerged from a non-additive entropy~\cite%
{ct} and that has broadly been applied \cite{application-kl}. The
comparisons presented are made in discrete time series that correspond to
the large majority of the time series available for analysis. The results
show that when the degree of complexity is small the two mutual information
measures studied provide the same answer. However, if we augment the
complexity of the signal, then we verify that the two non-linear measures
give different results.

\section{Theoretical preliminaries}

Consider Shannon entropy, $S$, as the average of the surprise, $s_{i}$,
associated with a system which has a certain probability distribution $%
\left\{ p_{i}^{\prime }\right\} $
\begin{equation}
S\equiv \sum_{i}p_{i}^{\prime }\,\ln \,\frac{1}{p_{i}^{\prime }}%
=\sum_{i}p_{i}^{\prime }\,s_{i}^{\prime }.
\end{equation}%
Suppose now that the system is modified or new measurements are made giving
rise to a new set, $\left\{ p_{i}\right\} $, of distributions associated
with the several states allowed by the system. From this new set, and for
each state $i$, we can define a new value for the surprise, $s_{i}=\ln \,%
\frac{1}{p_{i}}$, and its variation,
\begin{equation}
\Delta s_{i}=s_{i}^{\prime }-s_{i}.
\end{equation}%
Averaging $\Delta s_{i}$ with respect to distribution $\left\{ p_{i}\right\}
$ we have,
\begin{equation}
I\left( \left\{ p_{i}\right\} ,\left\{ p_{i}^{\prime }\right\} \right)
\equiv \sum_{i}p_{i}\,\Delta s_{i}=\sum_{i}p_{i}\,\ln \,\frac{p_{i}}{%
p_{i}^{\prime }},  \label{kullback-leibler}
\end{equation}%
which is the \emph{Kullback-Leibler entropy}. This measure has well-known
properties such as \textit{positiveness} and concaveness among many others~%
\cite{kullback}. Moreover, contrarily to $S$, $I\left( \left\{ p\right\}
,\left\{ p^{\prime }\right\} \right) $ is invariant under a change of
variables $x\rightarrow \tilde{x}=f\left( x\right) $, and is not symmetric
when we swap $p_{i}$ and $p_{i}^{\prime }$. The latter property invalidates
the possibility that $I\left( \left\{ p\right\} ,\left\{ p^{\prime }\right\}
\right) $ can be considered a metric distance. Nevertheless, we can still
use it as a distance measure in probability space.

Let us now consider that instead of using the surprise as we have just
defined, we have a $q$-surprise,
\begin{equation}
s_{i}^{\left( q\right) }=\ln _{q}\frac{1}{p_{i}},
\end{equation}%
where,
\begin{equation}
\ln _{q}\,x\equiv \frac{1-x^{1-q}}{1-q},  \label{q-ln}
\end{equation}%
(when $q\rightarrow 1$, $\ln _{q}\,x=\ln x$)~\cite{ct-bjp}. Therefore, the
variation of the $q$-surprise is
\begin{equation}
\Delta s_{i}^{\left( q\right) }\equiv s_{i}^{\left( q\right) \prime
}-s_{i}^{\left( q\right) }=\frac{\left( 1-\left[ p_{i}^{\prime }\right]
^{1-q}\right) -\left( 1-\left[ p_{i}\right] ^{1-q}\right) }{1-q}.
\end{equation}%
Computing the $q$-average of $\Delta s_{i}^{(q)}$ with respect to the
distribution $\left\{ p_{i}\right\} $ \cite{ct,emfc-ct},
\begin{equation}
E_{p}\left[ \Delta s_{i}^{(q)}\right] \equiv \sum_{i}\left[ p_{i}\right]
^{q}\,\Delta s_{i}^{(q)}=\sum_{i}\left[ p_{i}\right] ^{q}\,\frac{\left[ p_{i}%
\right] ^{1-q}-\left[ p_{i}^{\prime }\right] ^{1-q}}{1-q},
\label{q-media-kl}
\end{equation}%
and using Eq.~(\ref{q-ln}), we obtain the $q$\textit{-generalisation of
Kullback-Leibler entropy},
\begin{equation}
K_{q}\left( \left\{ p\right\} ,\left\{ p^{\prime }\right\} \right)
=-\sum_{i}p_{i}\,\ln _{q}\,\frac{p_{i}^{\prime }}{p_{i}},
\label{q-kullback-leibler}
\end{equation}%
for which $K_{1}\left( \left\{ p\right\} ,\left\{ p^{\prime }\right\}
\right) =I\left( \left\{ p\right\} ,\left\{ p^{\prime }\right\} \right) $.
Entropy $K_{q}\left( \left\{ p\right\} ,\left\{ p^{\prime }\right\} \right) $
is positive for $q>0$, negative for $q<0$, and null for $q=0$ or $%
p_{i}^{\prime }=p_{i}$ $\left( \forall _{i},q\right) $. It is also provable
that $K_{q}\left( \left\{ p\right\} ,\left\{ p^{\prime }\right\} \right) $
is concave for $q>0$ and convex for $q<0$ (other properties can be found in
Ref.~\cite{portesi}).

\paragraph*{$K_{q}$\textit{\ as a measure of dependence }\protect\cite%
{ct-kl-1998} -}

\label{dependence}

We shall now consider a bidimensional variable $z=\left( x,y\right) $ for
which we want to quantify the degree of dependence between $x$ and $y$. In
the application of $K_{q}$ to the analysis of the scale of dependence, the
most plausible distribution to be considered as the reference distribution
is the product of the marginal distributions,
\begin{equation}
p^{\prime }\left( x,y\right) =p_{1}\left( x\right) p_{2}\left( y\right) ,
\label{p1p2}
\end{equation}%
where
\begin{equation}
\begin{array}{c}
p_{1}\left( x\right) =\sum_{y}\,p\left( x,y\right) \\
p_{2}\left( y\right) =\sum_{x}\,p\left( x,y\right)%
\end{array}%
\end{equation}%
and $p\left( x,y\right) $ is the joint probability distribution.

Using Eq. (\ref{kullback-leibler}) we can verify that the Kullback-Leibler
entropy, which for this case is named as mutual information, can be written
as,
\begin{equation}
\begin{array}{ccc}
I\left( x,y\right) & = & S\left( x\right) +S\left( y\right) -S\left(
x,y\right) , \\
& = & S\left( x\right) -S\left( x|y\right) , \\
& = & S\left( y\right) -S\left( y|x\right) .%
\end{array}%
\end{equation}%
From the first equation it is simple to see that $I\left( x,y\right) $ only
becomes equal to zero when the variables $x$ and $y$ are independent,\textit{%
\ i.e.}, $p\left( x,y\right) =p_{1}\left( x\right) p_{2}\left( y\right) $.
Both of $S\left( x\right) $ and $S\left( y\right) $ refer to the entropies
of the respective marginal distributions and the entropy $S\left( x,y\right)
$ renders the entropy of the join distribution. Entropies like $S\left(
x|y\right) $ are computed as
\begin{equation}
S\left( x|y\right) =-\sum\limits_{x,y}p\left( x,y\right) \,\ln p\left(
x|y\right) \equiv -E_{p\left( x,y\right) }\left[ \ln p\left( x|y\right) %
\right] ,
\end{equation}%
in which $E_{\Pi }\left[ Y\right] $ represents the average of $Y$ associated
with distribution $\Pi $.

Considering Eq. (\ref{p1p2}), the $q$-generalisation, $K_{q}\left(
x,y\right) $, which is now called generalised mutual information, can be
expressed as,
\begin{equation}
K_{q}\left( x,y\right) =\sum_{x,y}\frac{\left[ p\left( x,y\right) \right]
^{q}}{1-q}\left\{ 1-\left[ p_{_{1}}\left( x\right) p_{2}\left( y\right) %
\right] ^{1-q}\right\} -\left\{ 1-\left[ p\left( x,y\right) \right]
^{1-q}\right\} ,  \label{q-kl-2d}
\end{equation}%
or,%
\begin{equation}
K_{q}\left( x,y\right) =-E_{p\left( x,y\right) }^{q}\left[ \ln
_{q}p_{1}\left( x\right) +\ln _{q}p_{2}\left( y\right) +\left( 1-q\right)
\ln _{q}p_{1}\left( x\right) \ln _{q}p_{2}\left( y\right) -\ln _{q}p\left(
x,y\right) \right] .  \label{i-q-1}
\end{equation}%
Writing,
\begin{equation}
p\left( x,y\right) =p_{1}\left( x\right) \tilde{p}\left( y|x\right) ,
\end{equation}%
and after some algebra, it is then possible to write Eq.~(\ref{i-q-1}) as,
\begin{equation}
K_{q}\left( x,y\right) =-E_{p\left( x,y\right) }^{q}\left[ \ln
_{q}\,p_{1}\left( x\right) -\ln _{q}\,\tilde{p}\left( y|x\right) -\left(
1-q\right) \left( \ln _{q}\,p_{1}\left( x\right) \ln _{q}\,\tilde{p}\left(
y|x\right) -\ln _{q}\,p_{1}\left( x\right) \ln _{q}\,p_{2}\left( y\right)
\right) \right] .  \label{i-q-2}
\end{equation}

From Eqs. (\ref{i-q-1}) and (\ref{i-q-2}), it is possible to determine the
maximum and the minimum values of $K_{q}\left( x,y\right) $. The minimum
value of $K_{q}\left( x,y\right) =0$, exactly corresponds to the case in
which $p\left( x,y\right) =p_{_{1}}\left( x\right) p_{2}\left( y\right) $.
Complementary, the maximum value occurs when there is a bi-univocal
dependence between the two variables, \textit{i.e.}, the maximum distance to
independence. In this case, the conditional entropy,
\begin{equation}
S_{q}^{(\bar{p}\left( x|y\right) )}=\sum\limits_{y}\left[ \bar{p}\left(
x|y\right) \right] ^{q}\ln _{q}\bar{p}\left( x|y\right) ,
\end{equation}%
must vanish since the uncertainty of having a value $x$ given $y$ is absent.
Analytically, this implies,
\begin{equation}
E_{p\left( x,y\right) }^{q}\left[ \ln _{q}\,\tilde{p}\left( y|x\right) %
\right] =E_{p\left( x,y\right) }^{q}\left[ \ln _{q}\,p_{1}\left( x\right)
\,\ln _{q}p_{2}\left( y\right) \right] =0.
\end{equation}%
This means that the maximum of $K_{q}\left( x,y\right) $ yields,
\begin{equation}
K_{q}^{MAX}\left( x,y\right) \equiv -E_{p\left( x,y\right) }^{q}\left[ \ln
_{q}p_{1}\left( x\right) +\left( 1-q\right) \ln _{q}p_{1}\left( x\right) \ln
_{q}p_{2}\left( y\right) \right] .
\end{equation}

The existence of upper and lower bounds allows us to define a ratio, $R_{q}$%
,
\begin{equation}
R_{q}=\frac{K_{q}}{K_{q}^{MAX}}\quad \in \left[ 0,1\right] ,  \label{q-ratio}
\end{equation}%
that defines the degree of dependence between the two variables $x$ and $y$.
For every case, there exists an optimal entropic index, $q^{op}$, which is
related to the degree of dependence, such that the gradient of $R_{q}$ is
more sensitive and therefore more capable of determining small variations in
the degree of dependence. In other words, $q^{op}$ is recognised as the
inflexion point of $R_{q}$ \textit{versus} $q$ curves. Regarding $q^{op}$
values, it is simple to verify that when $x$ and $y$ are independent $%
R_{q}=0 $ $\left( \forall _{q>0}\right) $ and optimal value is equal to
infinity, $q^{op}=\infty $. In the case of bi-univocal dependence, we have $%
R_{q}=1$ $\left( \forall _{q>0}\right) $, which implies in the limit of
total dependence that $q^{op}=0$. Thence, for a certain finite and positive
value of $q^{op}$, it is valid to ascribe a given degree of dependence
between the variables $x$ and $y$ that we are analysing .

To conclude this part let us briefly discuss a very specific case of
correlation in which the system tends to deflect from its past behaviour,
anti-correlation. Anti-correlation is easily verified in the space of
variables since the covariance provides to this case a negative value
yielding $C_{z}\left( t_{w},\tau \right) <0$. In the probability space,
i.e., when information measures are used, negative values cannot be obtained
(at least when $q>0$). In this case, it has been observed that value of $%
q^{op}$ presented by a anti-correlated time series is smaller than the value
presented by the same time series after shuffling \cite{internalreport}.
Therefore $q^{op}<q_{(shuffled)}^{op}$ can be taken as a signature of
anticorrelation.

\section{Application to time series analysis}

In what follows, we are going to apply the mutual information measures
described hereinabove to time series obtained from mathematical models and a
heuristic time series as well. Our goal is to determine the time scale, $T$,
at which each method considers the elements of a time series $x\left(
t\right) $ and $y\left( t\right) =x\left( t+\tau \right) $ as independent
from each other. Since we are dealing with finite time series some of the
analytical results we have previously presented are no longer valid. For
example, the value of total independence that is measured from a finite time
series is not $q^{op}=\infty $, but some finite value of $q^{op}$ instead.
However, for a specific time series, the level of independence can be
assessed by shuffling its elements in such a way that the existent
dependencies are wiped out. The scale of interest, $T_{K}$, is achieved when
$q^{op}\left( \tau \right) $ reaches the value of $q^{op}$ of a independent
shuffled series~\cite{goyo-smdq}. The same shuffling procedure allows us to
determine the noise level of the correlation function and the minimum value
of the mutual information $I$. The minimum concurs (within error margins) to
the mutual information of a shuffled series and this match give us the
respective time scale of interest $T_{I}$. The linear correlation scale of
independence, $T_{C}$, is obtained from the intersection of the correlation
function with the noise level, similarly to what is currently made in the
recurrence plot analysis technique~\cite{physreprecplot}. For sake of
simplicity, we are going to consider processes in stationary state whose
results are independent of the waiting time.

\subsection{Logistic map in the fully chaotic regime}

Consider the following non-linear dissipative map,%
\begin{equation}
x_{t+1}=1-2\,x_{t}^{2},\qquad x\in \left[ -1,1\right] ,  \label{logistic}
\end{equation}%
which corresponds to the logistic map in the fully chaotic regime. Equation (%
\ref{logistic}) is probably the most studied non-linear dynamical system
\cite{ott}. Elements of a time series obtained from iterating Eq.~(\ref%
{logistic}) are associated with the probability distribution,%
\begin{equation}
P\left( x\right) =\frac{1}{2\,B\left( \frac{1}{2},\frac{1}{2}\right) }\left(
\frac{1-x}{2}\right) ^{-\frac{1}{2}}\left( \frac{1+x}{2}\right) ^{-\frac{1}{2%
}},
\end{equation}%
with $B\left( \frac{1}{2},\frac{1}{2}\right) $ being the Beta function.
Furthermore, it can be shown that the summation, $\xi
_{N}=\sum_{i=1}^{N}x_{i}$ approaches the Gaussian distribution as $%
N\rightarrow \infty $ \cite{beck}.

\begin{figure}[tbh]
\begin{center}
\includegraphics[width=0.45\columnwidth,angle=0]{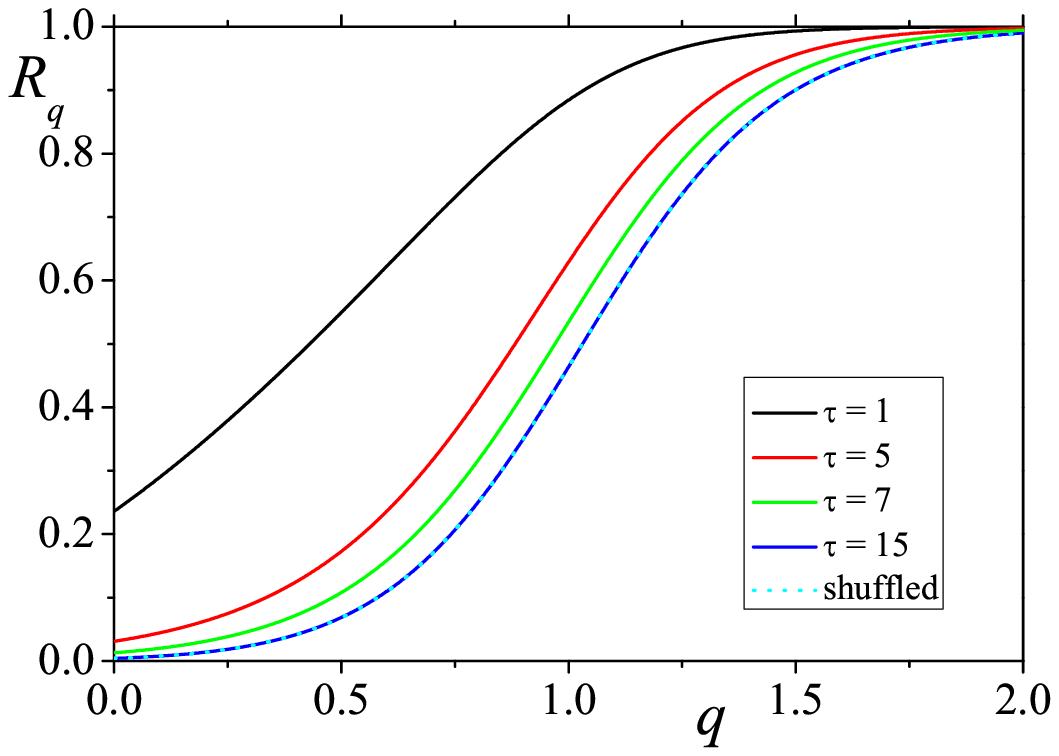} %
\includegraphics[width=0.45\columnwidth,angle=0]{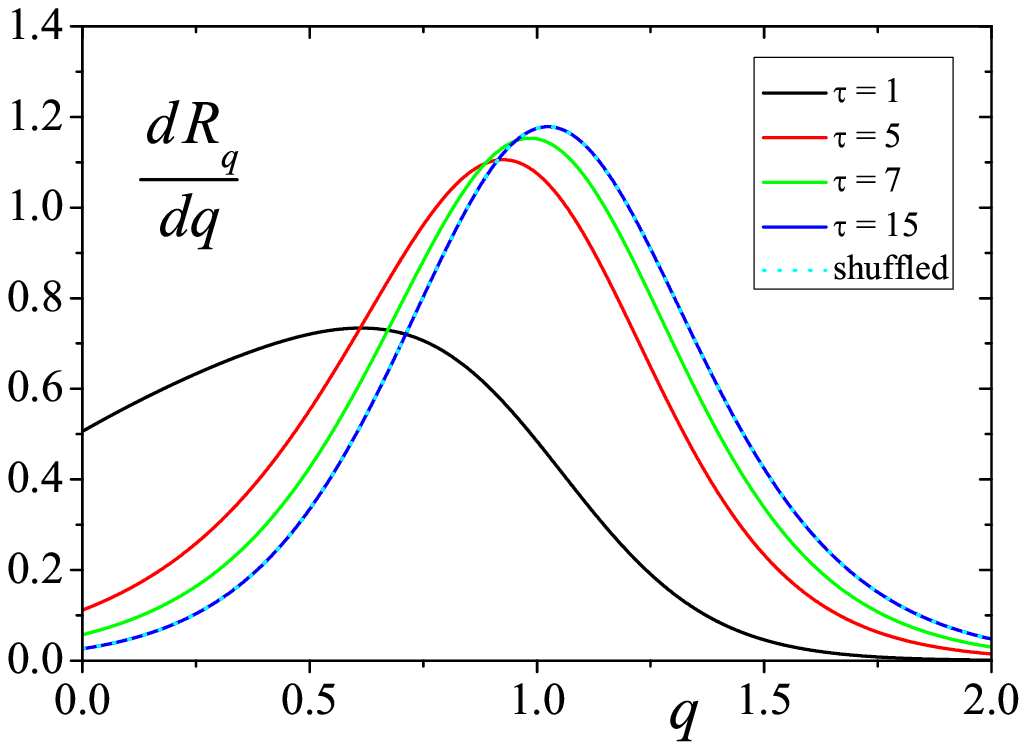}
\end{center}
\caption{\textit{Left:} Normalised generalisation of mutual information $%
R_{q}$ of $\left( x_{t},x_{t+\protect\tau }\right) $ \textit{vs.} $q$ for
series obtained from Eq.~(\protect\ref{logistic}) for several values of $%
\protect\tau $ and for series obtained after shuffling the elements from
logistic map sequences. \textit{Right:} Derivative of the curves in the left
panel with respect to $q$ \textit{vs.} $q$. The maxima correspond to the
inflexion points of $R\left( q\right) $, $q^{op}$, which are represented in
Fig. \protect\ref{fig-log}.}
\label{fig-rq}
\end{figure}

\begin{figure}[tbh]
\begin{center}
\includegraphics[width=0.45\columnwidth,angle=0]{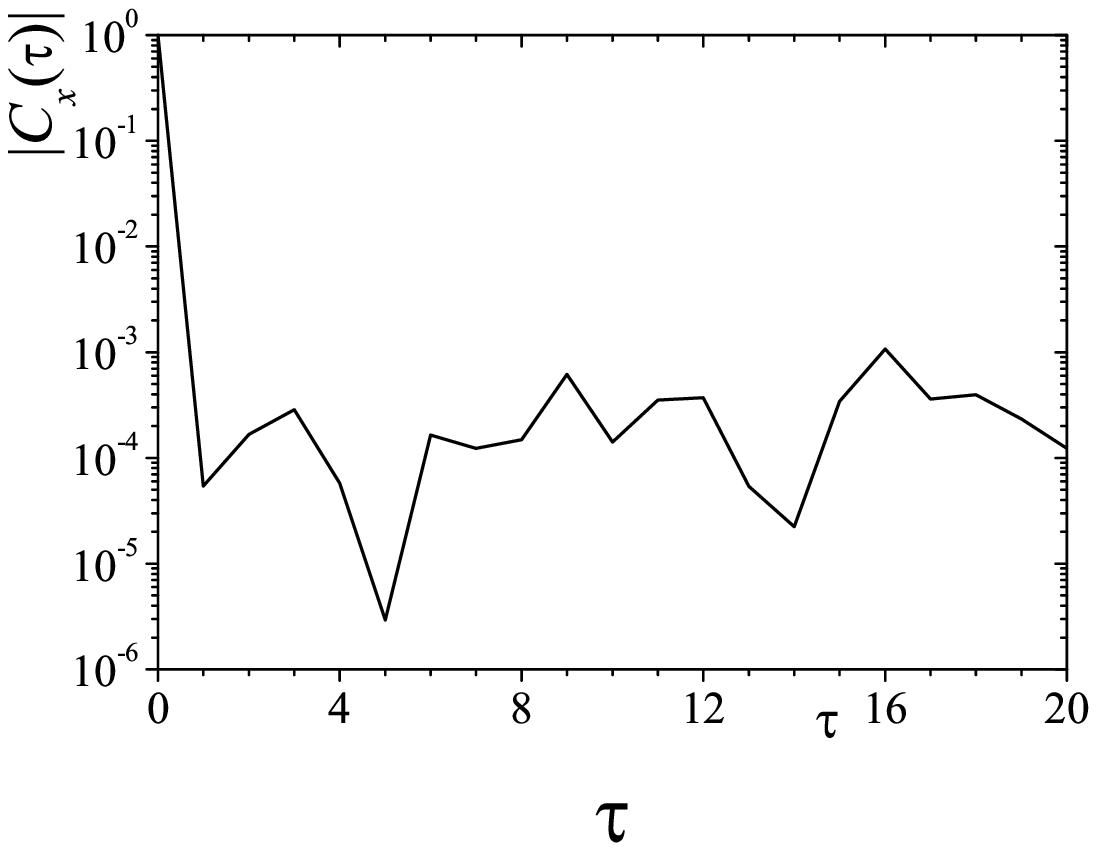} %
\includegraphics[width=0.45\columnwidth,angle=0]{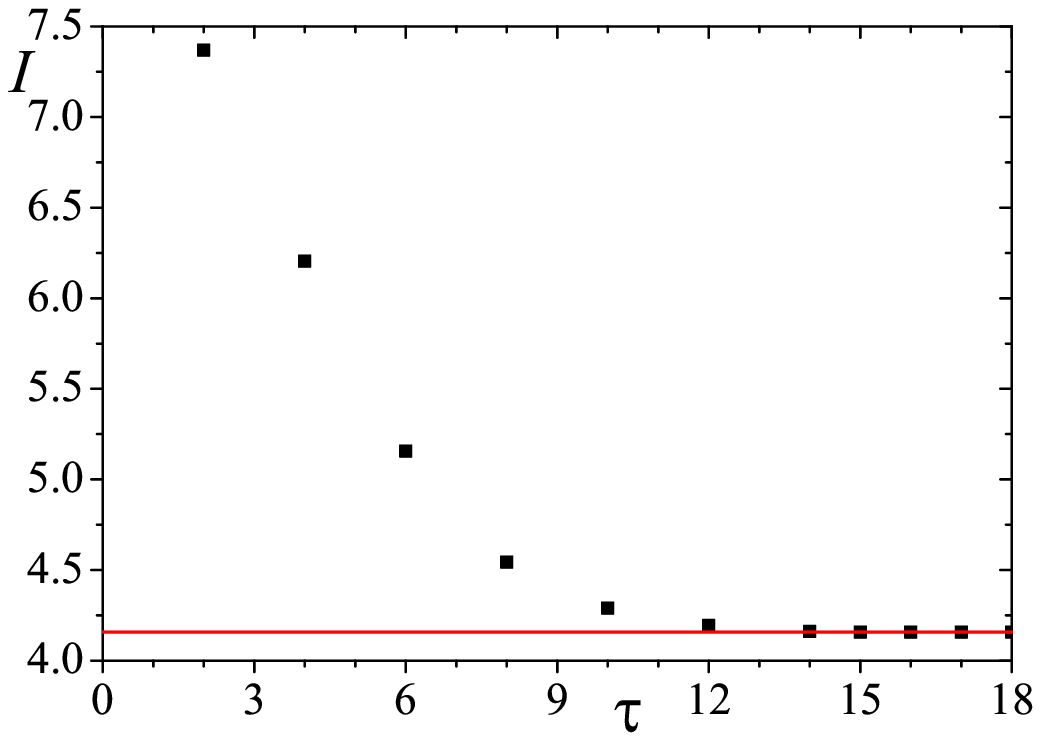} %
\includegraphics[width=0.45\columnwidth,angle=0]{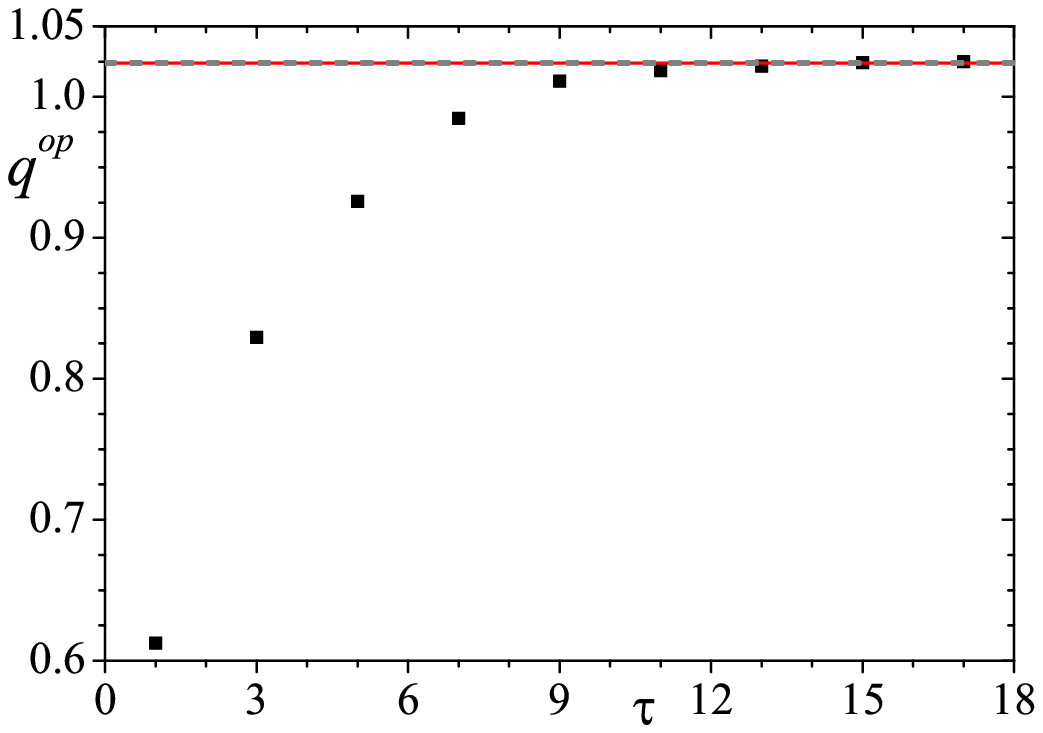}
\end{center}
\caption{\textit{Upper left:} Correlation function of the logistic map
\textit{versus} lag, $x_{t+1}=1-2\,x_{t}^{2}$. The correlation function is
at the noise level for $\protect\tau \geq 1$. \textit{Upper Right:} Mutual
information of the logistic map (points) and mutual information of logistic
map shuffled series (line). The matching occurs at the scale $T_{I}=15$.
\textit{Lower:} Optimal index \textit{versus} lag. The points have been
obtained from logistic map time series, the line represents noise level of $%
q^{op}$, and the grey lines represent the upper and lower bounds of error
margins. Once more, the match happens at the scale $T_{K}=15$. For every
case, averages over time series of $10^{6}$ elements are made. }
\label{fig-log}
\end{figure}

As expected, when we have analysed the autocorrelation function, we have
verified that $C_{x}\left( \tau \right) $ promptly attains the noise level.
As a matter of fact, we can write $C_{x}\left( \tau \right) =0$ $\left(
\forall \tau \geq 1\right) $, with higher-order correlations different from
zero as shown by Beck in \cite{beck1}. Computing mutual information $I$
between time series elements $x\left( t\right) $ and $x\left( t+\tau \right)
$, we have verified that the noise level is obtained for a lag $T_{I}=15$.
From the normalisation of the generalised mutual information measure, $%
R_{q}\left( \tau \right) $, and for each value of $\tau $, we have computed
the optimal values $q^{op}$. Comparing the values that were obtained from
the logistic map time series and the values obtained from the same time
series after shuffling their elements we have verified that the
characteristic time scale, for which the condition of independence between
variables prevails, is $T_{K}=15$. This scale is exactly the same time scale
indicated by the standard mutual information procedure. In Fig.~\ref{fig-rq}%
, we show typical curves of $R_{q}\left( \tau \right) $ for several values
of $\tau $. Each curve has been obtained from averages over different runs
(for specific values see caption in Fig. \ref{fig-log}). From the maximum of
every curve $\frac{dR_{q}}{dq}$ (right panel of Fig. \ref{fig-rq}) we have
computed $q^{op}\left( \tau \right) $ exhibited in Fig. \ref{fig-log}.

\subsection{Autoregressive conditional heteroskedastic process}

\begin{figure}[tbh]
\begin{center}
\includegraphics[width=0.45\columnwidth,angle=0]{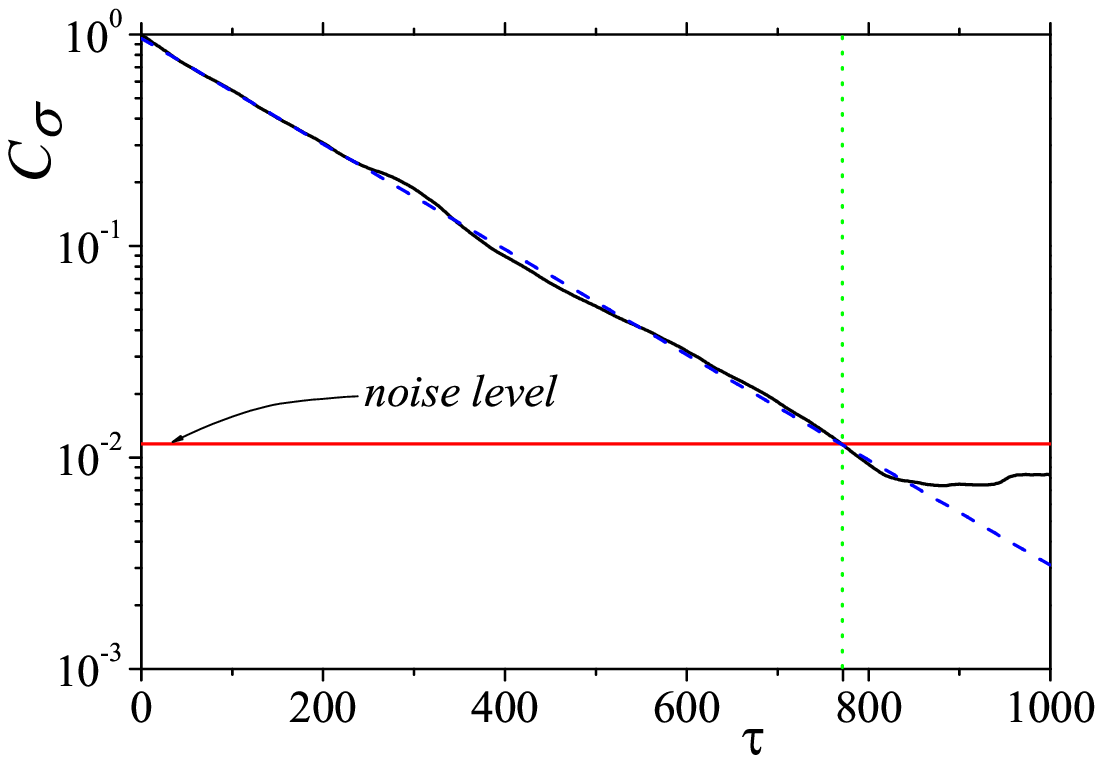} %
\includegraphics[width=0.45\columnwidth,angle=0]{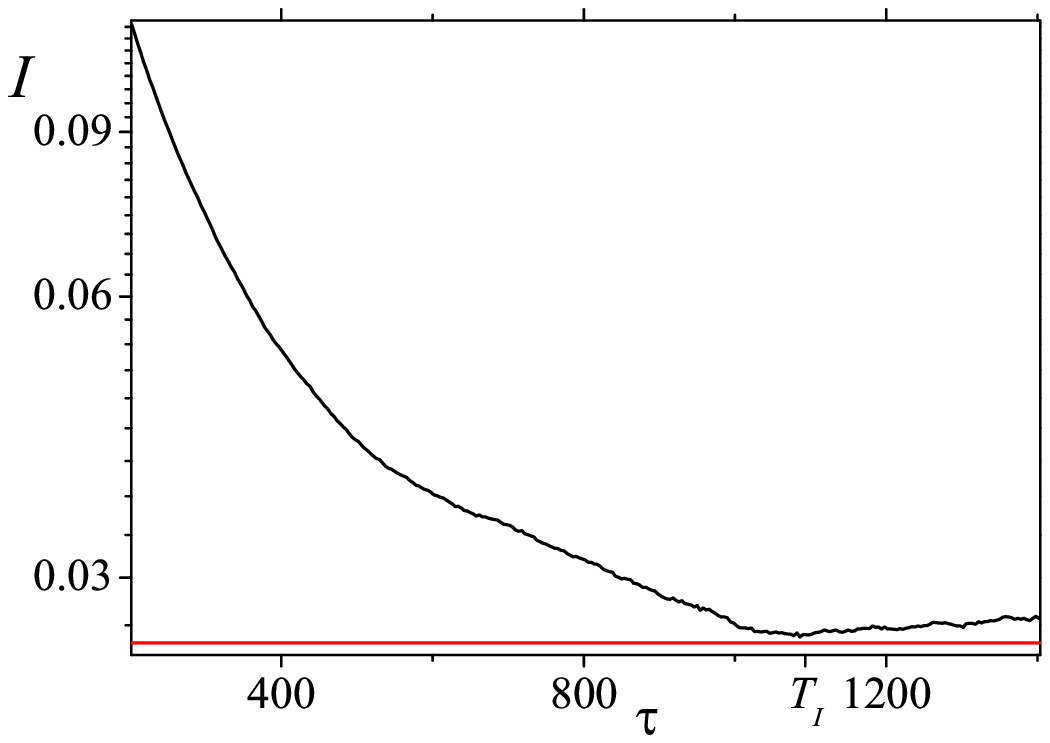} %
\includegraphics[width=0.45\columnwidth,angle=0]{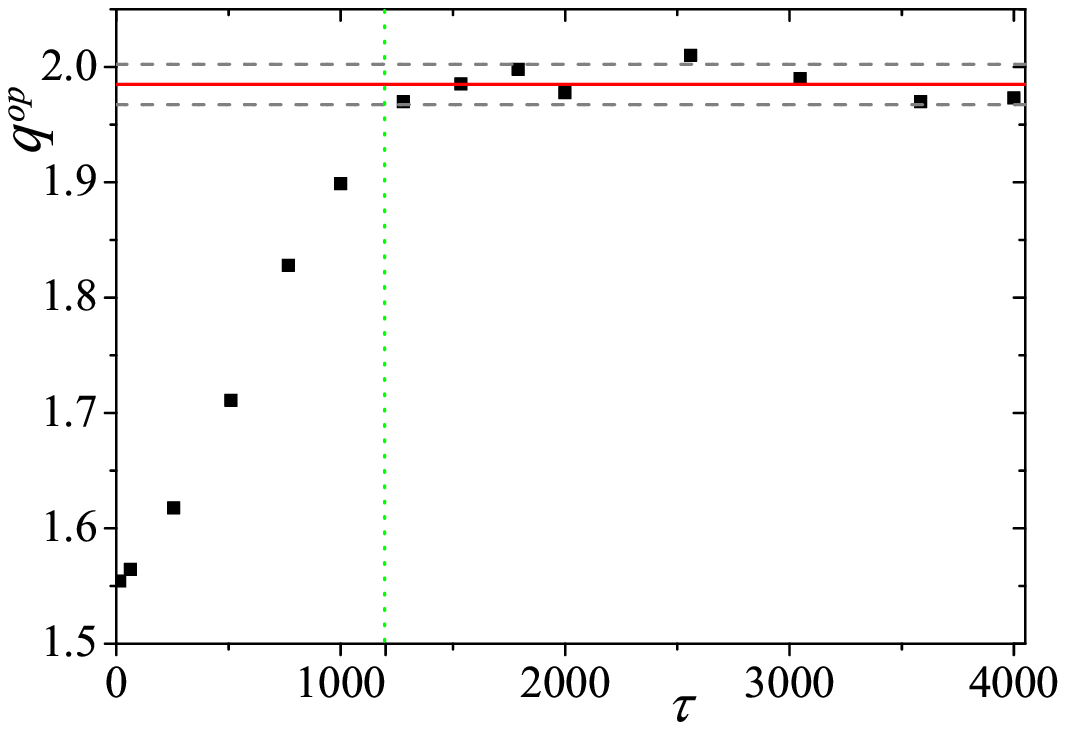}
\end{center}
\caption{\textit{Upper left:} Correlation function of a long-ranged
heteroskedastic process, Eq. \protect\ref{zefect}, \protect\cite{q-arch}
with $q_{m}=1$ [in $\log $-linear scale]. The correlation function is at the
noise level for $\protect\tau \geq T_{C}=772$ (dotted green vertical line).
The dashed blue line has a slope $400^{-1}$. \textit{Upper right:} Mutual
information of the same process (black line) and mutual information of
logistic map shuffled series (red line). The matching occurs at $T_{I}=1093$%
. \textit{Lower:} Optimal index versus lag. The points have been obtained
from logistic map time series and the red line represents noise level of $%
q^{op}$. The matching happens at $T_{K}\sim 1500$ clearly different from $%
T_{I}$. For every case, averages over time series of $10^{6}$ elements are
made after letting the process evolve for $10^{5}$ time steps to guarantee
stationarity. }
\label{fig-qarch}
\end{figure}

Many time series obtained from measurements in complex systems have shown
the peculiar feature of having (long-lasting) correlations in the magnitude
of its elements albeit their autocorrelation points to a white noise like
behaviour. Thus, the characterisation and modelling of the evolution of
instantaneous variance, $\sigma _{t}$, is of capital importance when we aim
to study that type of dynamics. To mimic this kind of time series, it has
been introduced by Engle the autoregressive conditional heteroskedastic
process (ARCH)~\cite{engle}. Here, we present an extreme case of a
generalisation that can be enclosed within the FIARCH class \cite%
{pier,q-arch}~\footnote{%
FIARCH stands for Fractionally Integrated ARCH}. Our variable is defined as,
\begin{equation}
x_{t}=\sigma _{t}\,\omega _{t},  \label{arch}
\end{equation}%
where $\omega _{t}$ is a stochastic variable usually associated with a
Gaussian distribution with null mean and unitary variance. The variable $%
\sigma _{t}$, also named as \textit{volatility} for historical reasons, is
defined as
\begin{equation}
\sigma _{t}^{2}=a+b\sum\limits_{i=t_{0}}^{t-1}\mathcal{K}\left( i-t+1\right)
\,x_{i}^{2},  \label{zefect}
\end{equation}%
where
\begin{equation}
\mathcal{K}\left( t^{\prime }\right) =\frac{1}{\mathcal{Z}\left( t^{\prime
}\right) }\exp \left[ \frac{t^{\prime }}{\varsigma }\right] ,\qquad \left(
t^{\prime }\leq 0,T>0\right)  \label{kernel}
\end{equation}%
with $\mathcal{Z}\left( t^{\prime }\right) $ being the normalisation. This
process originates non-Gaussian $x_{t}$ uncorrelated variables. Despite the
latter property, the autocorrelation function of $x_{t}^{2}$ (so as $\sigma
_{t}$ or $\left\vert x_{t}\right\vert $) presents an exponential decay. From
numerical implementation of Eqs. (\ref{arch})-(\ref{kernel}) with $a=0.5$, $%
b=0.99635$ (obtained for the case of price fluctuations studied in Ref. \cite%
{q-arch}), and $\varsigma =10$, we have obtained a set of time series from
which our results have been derived. To assure that the elements of the
analysed time series are in the stationary state, we have left each
numerical implementation run unrecorded for $10^{5}$ steps. As awaited, the
correlation function of $\sigma _{t}$ presents an exponential decay which
intersects the noise level at $\tau =T_{C}=772$. From our measurements of
the standard mutual information we have found a larger value of the
independence time which corresponds to a minimum at $T_{I}=1093$. Regarding
the application of the $q^{op}$ criterion, we have obtained an even larger
time to bear out independence between variables, $T_{K}\sim 1500$. This
value is clearly apart from $T_{I}$. Looking at the dashed (green) line in
the lower panel of Fig. \ref{fig-qarch} we see that the value of $%
q^{op}\left( T_{I}\right) $ is below the noise level (even considering error
margins). According to this criterion, this discrepancy points that at time
scale $T_{K}$ there is still a certain degree of dependence between
variables $\sigma _{t}$.

\subsection{Fluctuations of atmospheric temperature \label{temperature}}

\begin{figure}[tbp]
\begin{center}
\includegraphics[width=0.45\columnwidth,angle=0]{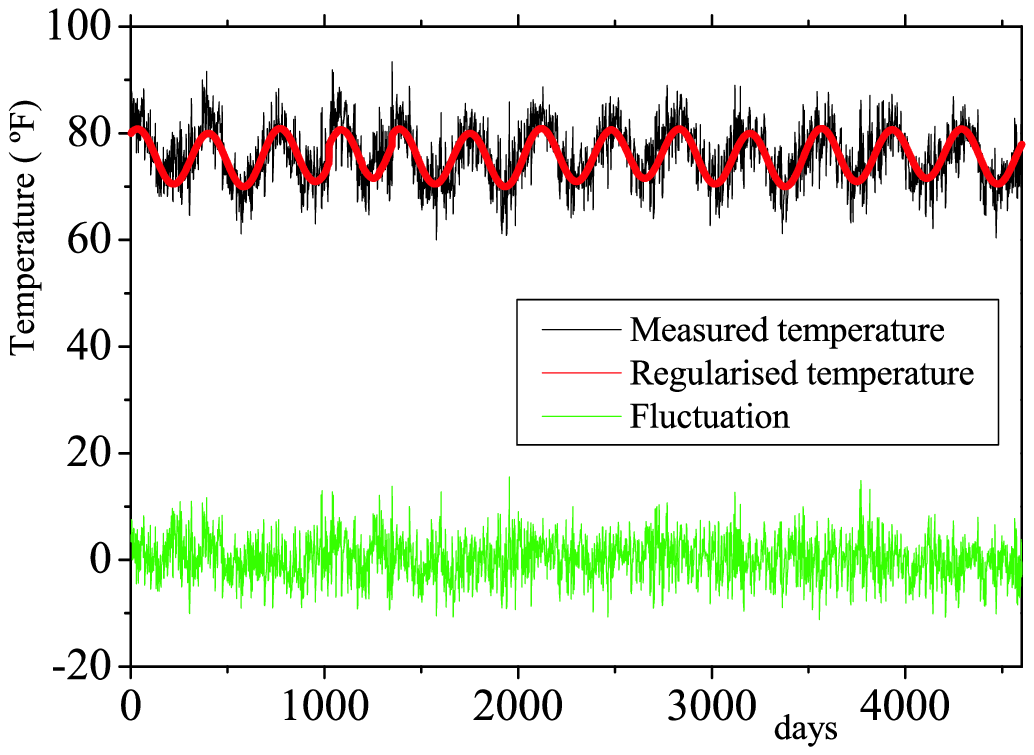} %
\includegraphics[width=0.45\columnwidth,angle=0]{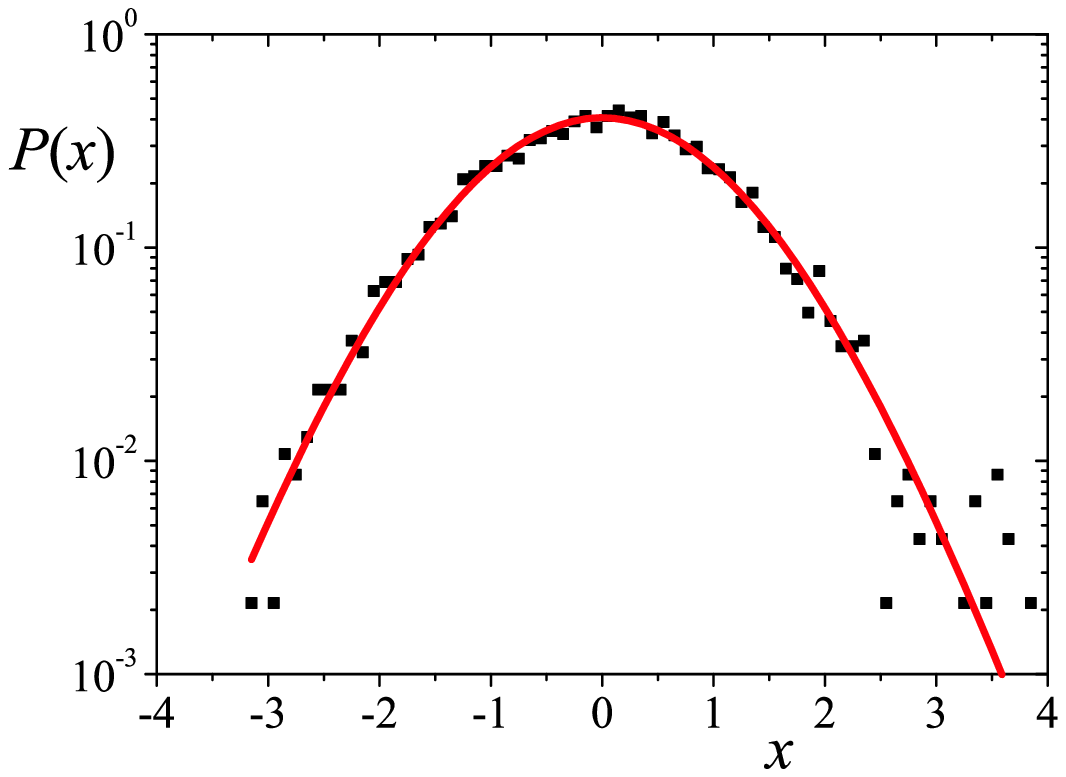}
\end{center}
\caption{\textit{Left:} Evolution of the atmospheric temperature (black
line), the regularised temperature (red line) and the fluctuation, $f$,
between measured and regularised temperatures (green line) at Rio de Janeiro
between the $1^{st}$ of January $1995$ and the $13^{th}$ of January $2008$
(temperatures in Fahrenheit degrees). \textit{Right: }Probability density
function $P\left( x\right) $ \textit{vs.} $x$ where $x$ represents the
detrended and normalised (by its standard deviation) temperature
fluctuations, $\left\langle f\right\rangle =0.15$ and $\protect\sigma %
_{f}=3.6$. As can be seen, $P\left( x\right) $ is very well fitted by a
Normal distribution with the error of adjustment being $\protect\chi %
^{2}=2.4\times 10^{-4}$ and $R^{2}=0.989$.}
\label{fig-series}
\end{figure}

\begin{figure}[tbp]
\begin{center}
\includegraphics[width=0.45\columnwidth,angle=0]{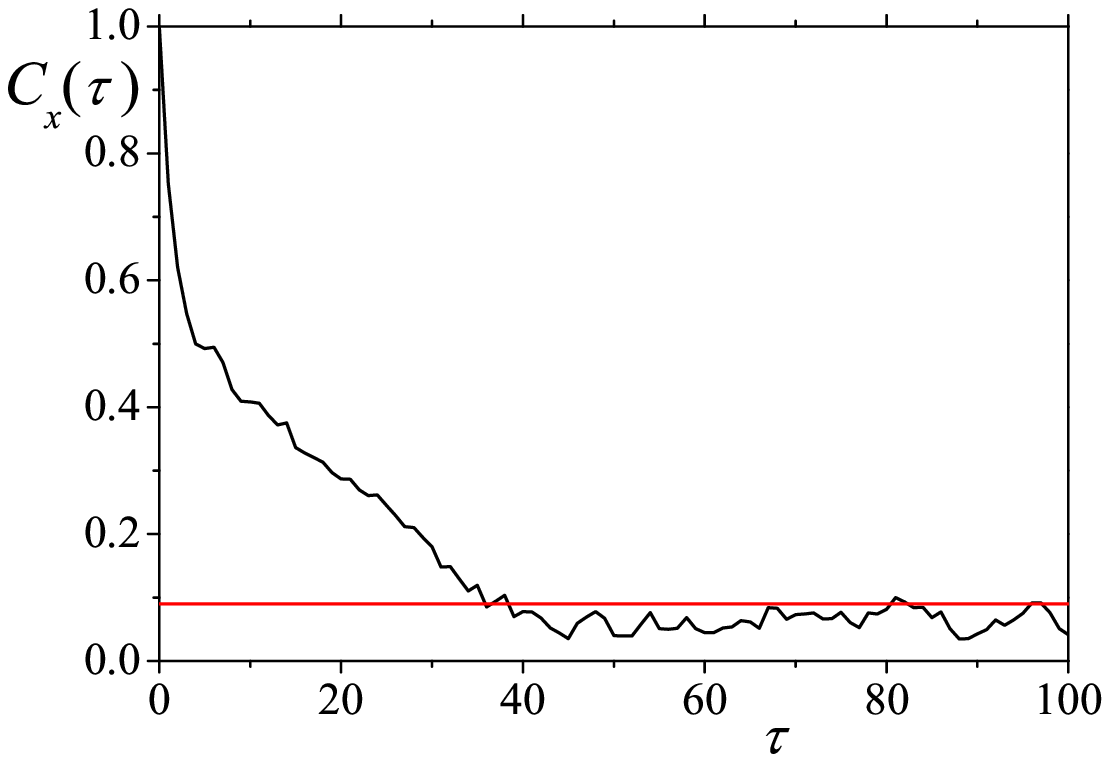} %
\includegraphics[width=0.45\columnwidth,angle=0]{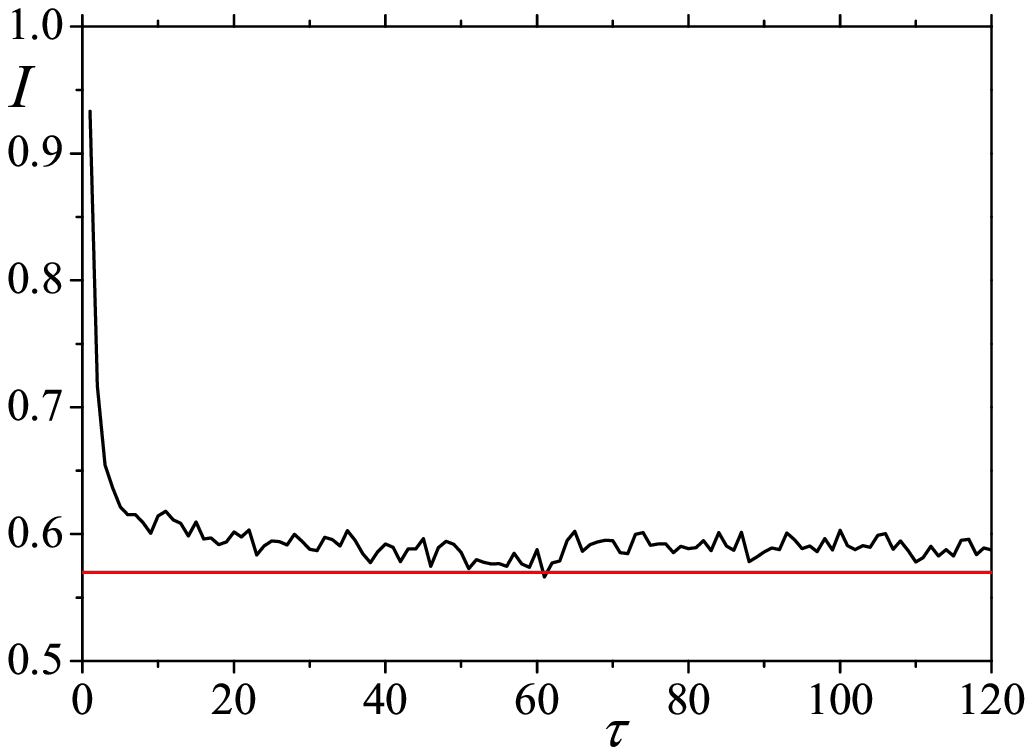} %
\includegraphics[width=0.45\columnwidth,angle=0]{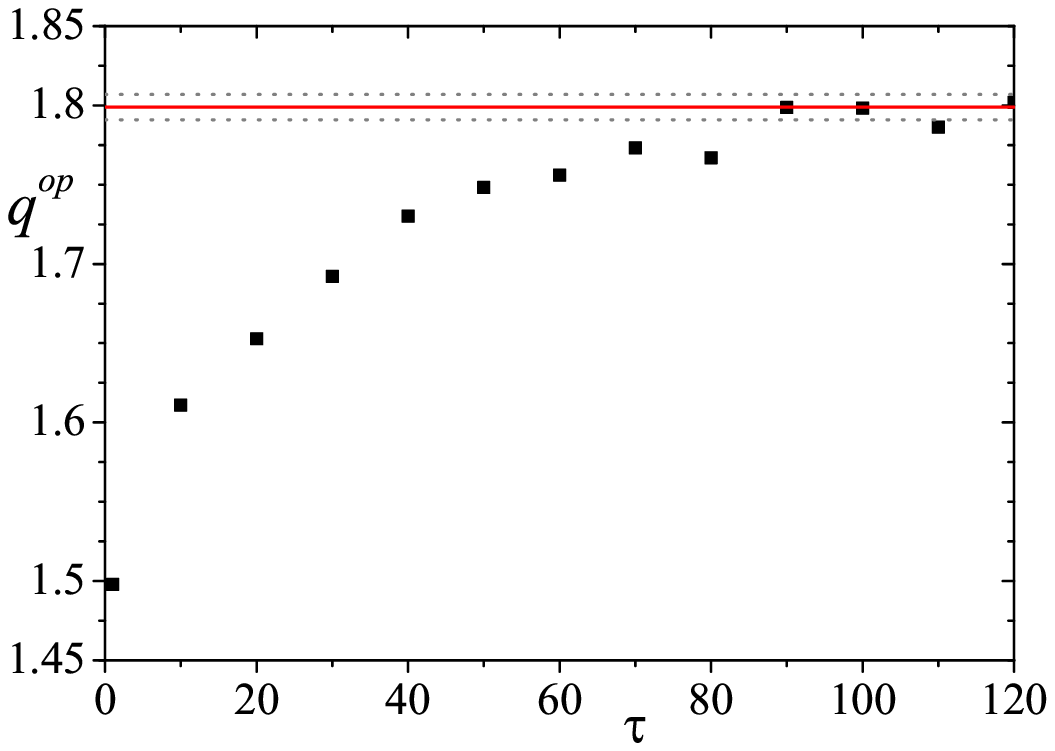}
\end{center}
\caption{\textit{Upper left:} Correlation function of the temperature
fluctuations presented in Fig. \protect\ref{fig-series}. The correlation
function attains the noise level at $\protect\tau =T_{C}=35$ days. \textit{%
Upper right:} Mutual information of the same series (black line)and mutual
information of shuffled series (red line). The matching occurs at $T_{I}=61$
days. \textit{Lower:} Optimal index versus lag. The points have been
obtained from the time series and red line represents the noise level of $%
q^{op}$. The equalisation happens at $T_{K}\sim 90$ days again plainly away
from $T_{I}$.}
\label{fig-temp}
\end{figure}

Fluctuations of atmospheric temperature have been intensively studied and a
paradigmatic case of time series analysis. In the next case, we analyse
fluctuations of the daily temperature with respect to the regularised
temperature in Rio de Janeiro (Brazil) between the $1^{st}$ of January $1995$
and the of $13^{th}$ of January $2008$ in a total of $4635$ observations~%
\cite{data}. Specifically, from the original time series we have obtained
the regularised temperature according to a standard procedure used in
climatology. The fluctuations have then been computed by finding the
difference between the measured temperature and the regularised temperature
(see Fig. \ref{fig-series} left panel). Computing the PDF of these
fluctuations we have found that they are very well described by a Gaussian
as shown in Fig. \ref{fig-series} right panel. In defiance of such a
Gaussian behaviour, when we have estimated the independence scale, we have
verified that the dynamics is in fact governed by long-memory effects~%
\footnote{%
The manisfestation of a Gaussian behaviour is by no means incompatible with
the existence of long-range memory as it can be understood from Refs.~\cite%
{q-arch,elephant}}. Numerically, we have noticed that the correlation
function comes into the noise level for $T_{C}=35$ days. Using the standard
mutual information we have obtained a minimum value for $T_{I}=61$ days
which indicates the existence of non-linearities governing the dynamics of
temperature fluctuations. Nevertheless, the scale given by $T_{I}$ appears
to be an intermediate one, like it has happened in the previous example,
since from the application of the $q^{op}$ we have obtained a larger upper
bound for dependence $T_{K}\approx 91$ days. We plot these results in Fig. %
\ref{fig-temp}. Again, we verify a hierarchical structure of independence
scales furnished by the correlation function, mutual information, and
generalised mutual information. This level of dependence might be related to
the fact that Rio de Janeiro is a onshore city, thus it is affected by the
stability provided at larger scales from the absorption or release of heat
by the sea.

\section{Final Remarks}

In this manuscript we have performed a comparative study between correlation
and dependence measures, namely the mutual information measure and a
generalised mutual information, defined within the context of non-additive
entropy $S_{q}$, aiming to obtain the respective independence scale between
elements. Our analysis has been performed on discrete time dynamical systems
with different levels of non-linearity and memory. Explicitly, we have
analysed the logistic map, a heteroskedastic process with long-lasting
memory and a natural time series namely the fluctuations of atmospheric
temperature. In the overall, our results have conveyed the well-known
capability of mutual information for determining the presence of
non-linearities. In addition, by means of increasing the memory of the
system, \textit{i.e.}, soaring the level of complexity, the differences
between the scale provided by mutual information, $I$, and by the criterion
based on $q^{op}$ come out with $T_{I}$ being consistently smaller than $%
T_{K}$. Hence, the comparison of the results given by different information
measures can be a helpful tool in order to opt for the most appropriate way
to model the dynamics related to the measurements of a certain observable or
to have an estimative about how further a forecast procedure can go
maintaining a sufficient level of reliability. We would also like to refer
that the work we have detailed points out the relevance of generalised
information measures like as it has been shown with the application of the
generalised-escort Tsallis entropy~\cite{generalisedescort} on the
distinction of pre-ital, ictal, and post-ictal stages of epileptic signals~%
\cite{epilepticsignal}. Last of all, we refer that this criterion to set
down the independence scale can also be used as an alternative method in the
determination of the independence scale and subsequent evaluation of the
embedding dimension of recurrence maps~\cite{internalreport,physreprecplot}.

\subsection*{Acknowledgements}

SMDQ aknowledges J.~de~Souza for performing the temperature regularisation
used in Sec.~\ref{temperature} and R.~Zillmer and G.~Savill for the comments made on the work
presented here above. This work benefited from financial support from
European Union through BRIDGET Project (MKTD-CD 2005029961).

\end{document}